# Improving alkaline fen functioning and *Liparis loeselii* (L.) Rich., 1817 preservation: towards a better water level management

Adrien Berquer[1*], Antoine Gazaix[2], Laura Czerniak[1], Valentin Dromard[1], Guillaume Meire[1], Gaëtan Rivière[1]

[1]Conservatoire d'espaces naturels des Hauts-de-France, 4 avenue de l'étoile du Sud, F-80440 Boves, France.

[2]Tour du Valat, Research Institute for the Conservation of Mediterranean Wetlands, F-13200 Arles, France

*corresponding author: a.berquer@cen-hautsdefrance.org

# Abstract

Alkaline fens are known to be wetlands that host large quantities of carbon, but also a huge biodiversity. However, anthropic pressures are degrading the peatlands with drainage, and conversion to agriculture or urbanization. These pressures induce externalities like a large release of greenhouse gas (GHG) by peat mineralization, and a loss of biodiversity, since they host numerous protected or endangered species. One of them is the *Liparis loeselii*, a small orchid facing declines in Europe and for which conservation measures are taken. Nevertheless, if recent studies inferred some factors shaping its population dynamics, they are still not clearly understood, particularly in fen contexts. This study aims at disentangle the processes shaping a continental population of *L. loeselii* in the Somme valley, among factors related to the hydrology of the site and the elevation of the individuals. We used Bayesian generalized linear models to infer the seasonal water level effects on the presence of *L. loeselii*. Moreover, we showed that the elevation of *L. loeselii* population is around the mean water levels of winter, and that the occurrence can be promoted by water level, notably in summer. Conversely, the highest water levels reported seemed detrimental to the *L. loeselii* population, suggesting a probably negative effect of flooding acting on local dispersal. Finally, this study provides insights to take restoration measures of alkaline fens, like a better hydrological functioning and an optimal water management.

# Introduction

Peatlands account for the half of wetlands on Earth (Convention on Wetlands, 2021). Although hosting a huge part of biodiversity they are currently under critical pressure, due to their exploitation for agriculture and peat extractions, involving drainage and peat degradation (Joosten, 2016). Moreover, peat degradation contributes directly to climate change, since the greenhouse gases (GHG) emissions increase with the drainage contribution to peat mineralization, which is an irreversible process (Joosten, 2016; Leifeld et al., 2011). Rewetting peatlands is therefore a major goal of peatland restoration since it contributes to limit from short to long term the peat degradation, the GHG emission, and the loss of habitats and biodiversity (Jurasinski et al., 2020; Kreyling et al., 2021; Zak and McInnes, 2022).

Alkaline fens are commonly found in Northern Europe, at topographic depressions accumulating water surging from calcareous aquifers. In Northern France, they are mainly located along the valleys of different rivers (Aa, Authie, Avre, Escaut, Somme), at topographic depressions (Sacy peatlands), or along the Northern France and Belgium coast (Marescaux et al., 2021). They also ecosystem services as various as the regulation of hydrologic regimes (e.g. flooding limitation, water provision in dry seasons), water quality, biodiversity refugia, carbon storage, and commons like tourism, history or leisure (Bonn et al., 2016; Groot et al., 2006; McInnes, 2013). In this geographical range, a lot of peatlands are actually degraded by drainage, agriculture and sometimes by trophic enrichment (Marescaux et al., 2021) and managed for a high primary production providing ecosystem services (grazing, hay production, wood production).

The combination of high trophic conditions with the abandonment of agropastoral management have led to tall and woody vegetal species encroachment, increasing the competition with smaller and heliophilous species (Kotowski et al., 2013; Middleton et al., 2006; Sonnier et al., 2023). *Liparis loeselii* (L.) Rich., 1817, the fen orchid, is one of the species unfavoured by the encroachment dynamics. This orchid, although widely distributed in Europe and Northern America, undergoes a general population decline, leading its near threatened species status at the European level (Bilz, 2011). This species is divided into two subspecies. Since *Liparis loeselii* subsp. *ovata* Riddelsdell ex. Godf. colonizes dune ponds, on sand, the subsp. *loeselii* is mostly found on continental areas and grows on peat. Northern France still count relatively large populations of this species, and its two sub-species, but also several well-preserved habitats to host them (both dune pannes and transition mires), the responsibility to the species conservation is high. Consequently, a national action program was developed allowing a collection of knowledges, about the species, and suggesting ways to its preservation (Valentin et al., 2010; Van Landuyt et al., 2015).

The factors shaping the distribution, abundance, and reproduction of *L. loeselii* remain poorly understood although a larger literature is available since the last decades (Błońska et al., 2016; Destiné, 2000; Jabłońska et al., 2011; Megre et al., 2018; Rolfsmeier, 2007; Roze et al., 2014; Van Landuyt et al., 2015; Wheeler et al., 1998). The first difficulty to study the species is the stochastic processes governing populations dynamics (Valentin et al., 2010). First of all, *L. loeselii* is a species benefitting of habitat dynamics, and management reopening sandy or peaty wetlands that contribute to its colonization. On the other hand, a lack of dynamics, and a management not allowing landscape reopening and pioneer habitat can lead populations to extinction. If local factors contribute to these rates, the dynamics and the preservation of *L. loeselii* are more understood at a metapopulation level (Grootjans et al., 2017). Metapopulations have been studied in dune pannes in Wadden sea, identifying that water levels and soil and water characteristics, as well as the time from the formation of the pond, contribute to the colonization/extinction of the species (Grootjans et al., 2017). This study testifies of different metapopulation levels, from local (adjacent pond to pond) to large (ponds of different islands) colonization dynamics. Water levels have also been shown to have an effect on the populations, the latter being more or less adapted to tolerate flooding (Grootjans et al., 2017). Changes in water chemistry are important, with an optimum spanning from desalination to decalcification (Grootjans et al., 2017; pers. obs., 2023). Indeed, Megre et al. (2018), evidenced that this tolerance is linked to functional traits, suggesting morphological adaptations -leaf shape and angle differed according to water levels-, and to fitness – individuals with higher aerenchymes in roots being less sensitive to long-term flooding. This study also underlined the ability of colonization by seeds measuring seed viability, concluding that partially water saturated substrates are favourable conditions for habitats to host *L. loeselii* populations. Moreover, a study within a population showed differed leaf and flower growth on individuals, according to the timing of snow melting, and therefore water levels, suggesting a plasticity to face environmental conditions and increase fitness while increasing the probability of reproduction and dispersal (Roze et al., 2014).

In the present study, we aim at understanding the effect of water level dynamics on the distribution of a population of *L. loeselii*. Our first hypothesis (i) the *L. loeselii* population is sharply distributed around the elevation mean of the fen and (ii) the population size may vary between years, and that high water levels in winter may promote the flowering of the species since it contributes to keep moist conditions to the habitat. Conversely, too long flooding could be detrimental to germination, growth and flowering depending of the flood timing. Thus we suggest that too high water levels in spring can reduce *L. loeselii* flowering and survival, but that too low water levels in this season may not induce germination, increasing stress conditions under drought, and promoting reproductive mechanisms.

These results should allow to know the factors contributing to the survival of the population. In a perspective of habitat restoration, the results would give insights to determine the hydrological management of the site, targeting optimal water levels, but also to determine the depth of a top-soil removal action to allow colonization of new habitats by *L. loeselii* and other transition mire species. This study aims at being an example of restoration planning meeting the preservation of both biodiversity and ecosystem functioning. Finally, this study aims at contributing to the knowledge of the factors shaping *L. loeselii* continental populations, since it remains poorly understood compared to littoral populations.

# Materiel and Methods

## Study Site

The study site is located in the Somme valley in Northern France, which is a valley remarkable by its huge area of peatland habitats -around 15 kha of minerotrophic alkaline rich fens (François, 2021). The site 'Grand marais de la queue' in Blangy-Tronville (Somme district, France; Figure 1), is known to host one of the three currently known continental station of *Liparis loeselii* (L.) Rich., 1817 (Meire and Rivière, 2019) in Northern France. In this site, *L. loeselii* subsp. *loeselii* is found on transition mires, that cover a surface of 8600 m². The first individual was found in 1999 and the population rapidly spread to adjacent transition mires. This mire is annually mowed with exportation, and is surrounded by woody habitats (Sphagno-Alnion glutinosae), that could contribute to encroachment and engage competition with rare species, *L. loeselii* among them (Meire and Rivière, 2019).

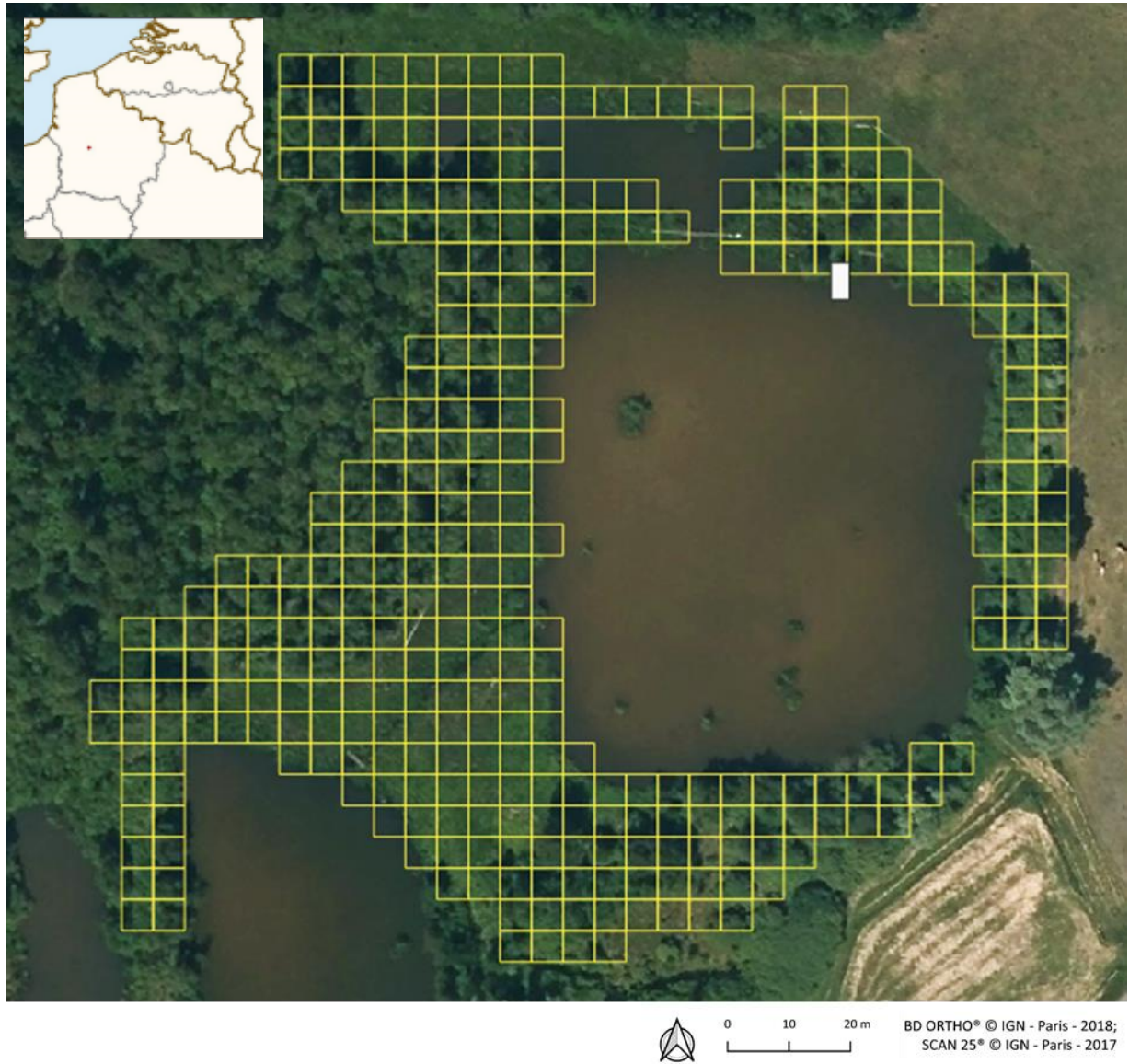

*Figure 1: Location of the L. loeselii population. The area is divided in squared 5x5 meters quadrats. The upper left map indicates (red dot) the location of Blangy-Tronville within the Hauts-de-France region. The white rectangle in the pond represents the location of the limnometric scale.*

# Data collection
## Water level monitoring

Water levels were measured once to twice a month on a limnometric scale located in the pond (Figure 1). This pond is an ancient peat extraction site, and is colonized around by a transition mire vegetation. The elevation of the transition mires within a year as well as water levels in micro-ponds located in the *L. loeselii* zone do not vary more than the GPS accuracy, as measured by the protocol of Berquer (2023). Indeed, it is rooted into the accumulated peat below, and therefore undergoes punctual flooding, during high water levels. We then consider uniform the water level on the current area of *L. loeselii* distribution on the site (Figure S1.A), although pond water level is generally from 1 to 4 cm above the piezometer level (Figure S1.B). We then calculated for each season (Winter, Spring, Summer and Fall), the mean, maximal and minimal water levels of the pond (Figure 2).

## Population monitoring

The distribution of *L. loeselii* was monitored using a grid of 5 × 5m square, covering all the potential area of the species (Figure 1). For each year, presence/absence of the species was defined cell by cell, considering overlapping of GPS point taken during the monitoring of the population. Each cell which was not overlapping by any GPS point was considered as absence cell for the species, for the evaluated year. In 2021, presence-absence was evaluated but individuals were not counted.

We monitored the population size of *L. loeselii* annually since 2010 (excepted on 2016, and from 2020 to 2022). Since the area is relatively small, all the area was searched for individuals, which were locally flagged to avoid double counting using colored sticks. Then GPS points were taken for each individual or group of individuals (less than 5 meters around the GPS point). Flowering individuals were counted separately from vegetative ones.

## Topographical measures

In 2017 and 2023, the stage of each individual found (627 in 2017, and 30 in 2023) on the area was measured, according to Meire (2019), as well as their elevation using a differential GNSS measurement computed with the RTK technology (Spectra Geospatial, SP20), providing a real-time accuracy of 10 mm in horizontal and 15 mm in vertical. Within each quadrat, we calculated the mean elevation of the soil as well as its standard deviation, from the data available on the RGE-Alti® database (IGN, 2018), using a Lidar technology. We extracted a raster with an accuracy on the site lesser than 0.3 meters, converted into 1 m² pixels. This data treatment was performed on QGIS version 3.28.4 'Firenze'.

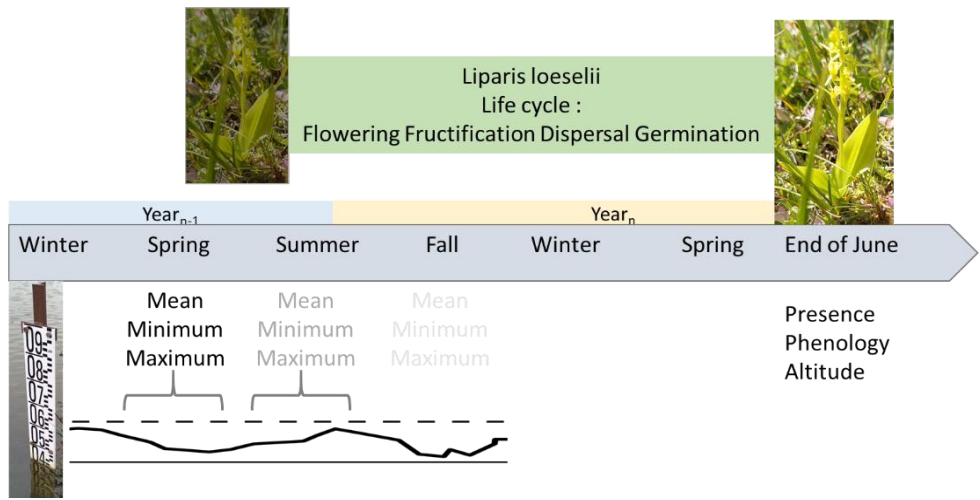

*Figure 2: Sampling strategy and timing along the seasons of the year including the processing of the water levels for each season (trimesters). Data monitoring of L. loeselii population differs according to the year.*

## Statistical analysis

We first modelled the occurrence of *L. loeselii* according to the water levels of the six different preceding seasons. For a growing season of *L. loeselii* in Summer of a given year (year$_n$), we took as variables the differences between the mean elevation quadrat and the water levels of Spring and Winter of the year$_n$, and the water levels of Fall, Summer, Spring, and Winter of the year$_{n-1}$. Indeed, the life cycle of *L. loeselii* includes an invisible protocorm stage that may occur several months before the vegetative stage (Valentin et al., 2010), a germination that has been documented by Tsutsumi et al. (2011) on three other *Liparis* species.

Each model was assessed using a Bayesian Generalized Linear Model (Muth et al., 2018), assuming a binomial distribution due to presence-absence data, with a Normal distribution as a prior. We used Monte-Carlo Markov Chains (MCMC), using two chains with 5,000 iterations. Concurrent models using a model simplification were tested, discarding the non-significant variables, as well as an improvement of the fit adding a quadratic term of each variable in other concurrent models. The evaluation of each model was based on parameters assessing the fitness of the model like the Rhat considering a threshold of 0.7, then we checked for convergence (function shinystan, Gabry et al., 2022). The selection of the final model was performed using a Leave-One-Out (LOO) procedure, giving with the EPLD and SE differences a comparison with concurrent models as detailed by Sivula et al. (2022).

In a second analysis performed on the observations and the elevation measured for each found individual in 2017 and 2023, we compared the altitudinal distribution of the flowering, vegetative and total population with the water levels, using Welch t-tests, to study if the elevation range of *L. loeselii* individuals undergoes regular drought (water level significantly below ground), or flooding (water level significantly above-ground).

All the statistical analyses were performed on R software (version 4.2.2, R Core Team, 2023), using the packages rstanarm (version 2.21.4, Stan Development Team, 2023) and shinystan (version 2.6.0, Gabry et al., 2022).

# Results

All the models presented showed adequate parameterization and convergence. If concurrent models were preferred in some cases, discarding one to four variables, quadratic effects never improved our models.

## Occurrence of *L. loeselii* individuals within quadrats

We performed Bayesian generalized linear models to explain the presence/absence variable by the elevation standard deviation, and by mean, maximal or minimal water levels measured below the quadrat mean ground elevation, in different seasons before the observations. We showed a significant negative effect of the elevational heterogeneity on the presence of individuals on the quadrat. The mean and minimal spring water levels had negative effects, whereas the maximal water level in spring had a positive effect on the occurrence. Conversely, maximal winter water level had a negative effect but the mean and minimal winter water levels had positive effects on the presence of *L. loeselii* in a quadrat. Maximal and minimal water levels in fall had positive effects, which was not the case for the mean water level. The mean and minimal water levels in summer had positive effects. High water levels (both, mean, minimal and maximal) during the winter of the preceding year had always significantly positive effects on the occurrence (Table 1).

Table 1: Results of the models of the presence of L. loeselii within quadrats according to elevation (mean and sd) and (a) mean, (b) maximal, and (c) minimal water levels.

| (a) | | | **Mean Water level** | | | | | |
|---|---|---|---|---|---|---|---|---|
| | Intercept | Sd elevation | Spring $Year_n$ | Winter $Year_n$ | Fall $Year_{n-1}$ | Summer $Year_{n-1}$ | Spring $Year_{n-1}$ | Winter $Year_{n-1}$ |
| Occurrence | 0.8 (±0.1) | -5.8 (±1.6) | -3.0 (±1.2) | 6.9 (±1.2) | ns | 2.9 (±0.8) | ns | 5.7 (±0.9) |
| (b) | | | **Maximal Water level** | | | | | |

|              | Intercept   | Sd elevation | Spring Year$_n$ | Winter Year$_n$ | Fall Year$_{n-1}$ | Summer Year$_{n-1}$ | Spring Year$_{n-1}$ | Winter Year$_{n-1}$ |
|---|---|---|---|---|---|---|---|---|
| Occurrence   | ns          | -6.9 (±1.6)  | 6.6 (±0.9)  | -1.6 (±0.9) | 2.1 (±0.7)  | ns          | ns          | 3.8 (±0.8)  |
| (c)          |             |              |             | **Minimal water level** |   |             |             |             |
|              | Intercept   | Sd elevation | Spring Year$_n$ | Winter Year$_n$ | Fall Year$_{n-1}$ | Summer Year$_{n-1}$ | Spring Year$_{n-1}$ | Winter Year$_{n-1}$ |
| Occurrence   | 2.2 (±0.2)  | -4.5 (±1.6)  | -4.7 (±0.8) | 8.1 (±0.9)  | 6.6 (±1.1)  | 2.3 (±1.2)  | -2.1 (±1.1) | 5.1 (±1.0)  |

For each year, we performed t-tests to highlight an elevation mean difference between quadrats with or without *L. loeselii* individuals. Every year, with no exception, the quadrats in which a presence of *L. loeselii* was reported were significantly at a lower elevation than the quadrats in which no individual was reported. The same pattern was found with the standard deviation in quadrat elevation, showing significantly higher heterogeneity of the elevation in absent quadrats than in presence quadrats. The mean elevation of quadrats with a presence of *L. loeselii* individuals remained between 24.58 and 24.61 meters above sea level, and the mean elevation of absence quadrats remained between 24.70 and 24.74 m.a.s.l.. In the same way presence quadrats had a standard deviation around 0.03 m whereas absence quadrats have a standard deviation around 0.05 m (Table S1).

## Altitudinal distribution of Liparis loeselii

The population of *L. loeselii* was counted from 2010 to 2023, at the exception of the years 2016, and 2020 to 2022. The population ranges from 30 individuals in 2023 to 1659 in 2015 (Figure 3). The proportion of flowering individuals ranges from 20% (2015) to 80% (2013).

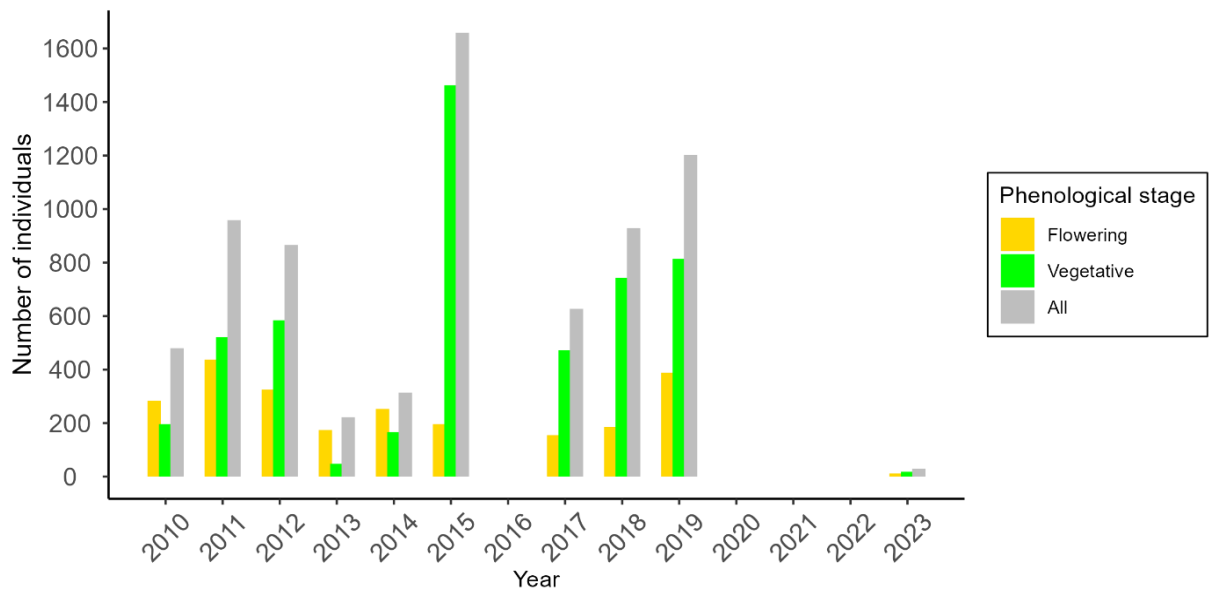

*Figure 3: Number of individuals of Liparis loeselii observed from 2010 to 2023. The count is represented according to the phenological stage (yellow, flowering; green, vegetative; grey for the total population). In 2016, 2020, 2021 and 2022, the individuals were not counted according to the same protocol, i.e. with less exhaustivity.*

We then compared the differences between the mean elevation of the population of *L. loeselii* of the whole site and for each phenological stage, with the mean elevation of water level, for each season, calculated over five years before the observations (i.e. from June 2012 to June 2017 for the comparison of the 2017 population, and from June 2018 to June 2023 for the comparison in 2023). We consider that a five-year time span may include both driest to wettest years and is more adapted to understand the fluctuations than a one-year time span. We used Welch two samples t-test since we checked for a normal distribution of each compared sample, with different variances (Table 1).

We did not find any difference between the mean elevation of *L. loeselii* individuals and the water levels in winter, whatever the phenological stage considered, meaning a water depth not significantly different from the *L. loeselii* ground. Conversely, we found a mean water level significantly below the *L. loeselii* altitudinal distribution in summer and fall. In spring we only found a significant difference on the vegetative and total population of 2017, that can be due to a fewer number of individuals in 2023 (respectively 627 and 30), meaning that the *L. loeselii* ground elevation undergoes at least punctual flooding some years. The mean water level in fall was systematically the lowest, whereas the highest mean water levels occurred in spring and winter.

*Table 1: Comparison of the elevation distribution of L. loeselii with the mean water level in different seasons. Results are split between 2017 (upper table) and 2023 (lower table), since the monitoring of the altitudinal distribution of L. loeselii population occured only during these two years.*

| Year | Stage | Season | Mean L. loeselii Elevation | Mean water level of the season | t-value | df | p-value | |
|---|---|---|---|---|---|---|---|---|
| 2017 | Flower N = 155 | Winter | 24.471 | 24.459 | 0.911 | 33 | 0.369 | |
| | | Spring | 24.471 | 24.475 | -0.323 | 36 | 0.748 | |
| | | Summer | 24.471 | 24.405 | 3.500 | 18 | **0.003** | ** |
| | | Fall | 24.471 | 24.391 | 7.437 | 29 | **<0.001** | *** |
| | Vegetative N = 472 | Winter | 24.444 | 24.459 | -1.305 | 27 | 0.203 | |
| | | Spring | 24.444 | 24.475 | -2.579 | 29 | **0.015** | * |
| | | Summer | 24.444 | 24.405 | 2.143 | 16 | **0.045** | * |
| | | Fall | 24.444 | 24.391 | 5.392 | 21 | **<0.001** | *** |
| | Total N = 627 | Winter | 24.449 | 24.459 | -0.954 | 26 | 0.349 | |
| | | Spring | 24.449 | 24.475 | -2.243 | 28 | **0.033** | * |
| | | Summer | 24.449 | 24.405 | 2.378 | 16 | **0.030** | * |
| | | Fall | 24.449 | 24.391 | 5.878 | 20 | **<0.001** | *** |
| 2023 | Flower N = 12 | Winter | 24.501 | 24.553 | -1.500 | 14 | 0.155 | |
| | | Spring | 24.501 | 24.549 | -1.599 | 17 | 0.128 | |
| | | Summer | 24.501 | 24.401 | 2.402 | 16 | **0.028** | * |
| | | Fall | 24.501 | 24.377 | 3.295 | 16 | **0.004** | ** |
| | Vegetative N = 18 | Winter | 24.513 | 24.553 | -1.164 | 13 | 0.265 | |
| | | Spring | 24.513 | 24.549 | -1.218 | 16 | 0.241 | |
| | | Summer | 24.513 | 24.401 | 2.747 | 15 | **0.015** | * |
| | | Fall | 24.513 | 24.377 | 3.672 | 15 | **0.002** | ** |
| | Total N = 30 | Winter | 24.509 | 24.553 | -1.309 | 13 | 0.213 | |
| | | Spring | 24.509 | 24.549 | -1.387 | 16 | 0.185 | |
| | | Summer | 24.509 | 24.401 | 2.629 | 15 | 0.019 | * |
| | | Fall | 24.509 | 24.377 | 3.542 | 15 | 0.003 | ** |

# Discussion Conclusion

Overall, we confirmed the expectation through the analysis of results. *L. loeselii*, in Blangy-Tronville, confirmed at having an irregular population dynamic, which is a general pattern for this species (Valentin et al., 2010). We also found effects of the water level on the probability of occurrence.

## Trend of the *Liparis loeselii* population

We showed with the monitoring of both flowering and vegetative individuals of *L. loeselii* in the site Grand Marais de la Queue, that the population undergoes large fluctuations between years (Figure 2). This dynamic was previously described in some studies (Grootjans et al., 2017; Hendoux et al., 2001; Valentin et al., 2010), but remains poorly understood at least on inner land populations. In the case we studied, the population fluctuates both in total, flowering and vegetative individuals. However, the years with the highest numbers of individuals were mainly the years with the highest percentage of vegetative individuals. Conversely, the years with a large proportion of flowering individuals are the years with the lowest total of individuals. This pattern may underline years particularly suitable for the *L. loeselii* reproduction or dispersal, or conversely unsuitable years promoting stress-induced flowering (Takeno, 2016). Since the life cycle of *L. loeselii* spans on several years (Valentin et al., 2010), we suggested that the fluctuation may be due to preceding seasons, including climatic conditions and site management. We observed that a high number of observations of vegetative individuals may follow a year with relatively better conditions for reproduction and recruitment. Conversely, in case of harsh conditions for *L. loeselii*, we could suggest that the effect on the following year, would be a negative effect on vegetative population, because a low recruitment, but an unchanged or increased number of flowering individuals, being vegetative the previous year, leading to a higher proportion of flowering individuals (Takeno, 2016; Tsutsumi et al., 2011). Regarding the vegetation management, we did not find any effect of the timing since the last mowing, though the trend is decreasing for vegetative individuals, but based on few observations (Figure S2).

## Effects of elevation and water levels on the *L. loeselii* occurrence

Peatland functioning and peat conservation are deeply dependant of the water levels, and the restoration action may also include a restoration of the hydrology functioning using weirs to maintain water levels particularly in driest summers that occur more frequently in Northern France with the global warming. To meet the hydrology restoration with the *L. loeselii* preservation, we performed models of presence/absence (i.e. occurrence) in a quadrat with the elevation heterogeneity and water levels as variables. We confirm that a heterogenous ground elevation had negative effects on the occurrence of an individual in a quadrat. Moreover, we found that spring water levels decrease the occurrence whereas in winter a higher minimal water level increased it. In summer, the water level had a positive effect. Therefore, retaining water on the site, and improving hydrology with restoration, seem to be of crucial importance for the *L. loeselii* preservation, but also to avoid a drought of the peat superficial layer that can contribute to peat degradation, soil mineralization, a greenhouse gas release, and subsidence (Hatala et al., 2012; Herrera-García et al., 2021; Ikkala et al., 2021). In fall, the increasing occurrence with high water levels may be due to a promoted dispersal on the nearby area, due to higher water level, compared to dryer years. The fall being the season with the lowest water levels, we may suggest that a drought could not disperse seeds or unfavour the development of the protocorm stage, i.e. the stage that allow some orchid species to winter. A lack of precipitation and increased temperature being expected with the global warming, this effect should have consequences for the *L. loeselii* preservation. Conversely, high water levels during the dispersal season might promote large scale dispersal. This process had previously been studied by Vanden Broeck et al. (2014) comparing the genetic variation of population located in dunes and fens across Europe. These effects were expected, meeting the results of Jabłońska et al. (2011) at the habitat level, the plant community hosting *L. loeselii* being altered by decreasing water levels. Our results also meet the effect of water levels found by Megre et al. (2018) on the survival rate of *L. loeselii*, hence affecting populations.

## Comparison between the elevation of *L. loeselii* and the water table

To show if a flood, even temporary, can disperse at short or large scale, and to show if a drought might be responsible to unfavourable conditions for an optimal growth, germination or survival, we performed tests comparing the elevation of the individuals with the elevation of the water levels. We found that in winter and spring, the water level rises around the mean elevation of *L. loeselii* individuals, and eventually flood the area, which is not the case in summer and fall when the water level is significantly below the *L. loeselii* ground. Though we found differences in mean elevation between 2017 and 2023, years where we gathered data with the differential GNSS, the differences between the *L. loeselii* altitudinal distribution and the water levels were the same within each year. Thanks to these results, combined with the ones from the models, we suggest that the *L. loeselii* dispersal in summer requires water to disperse, but it also depends of low water levels to allow the propagules to beach in a suitable nearby area, rather than being an effect due to flooding. The mean water levels in winter and spring preceding 2023 are higher than those of 2017, whereas the ones in summer and fall do not. These conditions, with the results of the model on the occurrence might explain a recent decrease of occurrence the recent year. Although the *L. loeselii*, an hygrophile species, is promoted by constant wet conditions, highest maximum water levels, suggesting immersion, or temporary flooding, might be detrimental (Megre et al., 2018). Considering management implications, it should be better for the *L. loeselii* conservation to maintain or increase a water release in winter or spring but keep high water levels at the end of spring to anticipate the prejudicial summer drought.

## A strait range of elevation is optimal for the presence of *L. loeselii*

We found that in the area that we considered suitable for the presence of *L. loeselii*, the quadrats where an actual presence of *L. loeselii* was reported were significantly lower than quadrats where no individuals were reported. In the latter, the higher elevation might avoid the optimal and constant wet conditions required by *L. loeselii*. Although the Somme stream often receive nutrients from agriculture and urbanization and often show unsuitable water to keep oligotrophic conditions in the peatlands of the valley, analysis of the water composition of Blangy-Tronville show a water quality compatible with favourable conditions for the development of protected or threatened habitats or species (pH = 7.86, Nitrate ≤ 0.5 mg.L$^{-1}$, Nitrite ≤ 0.05 mg.L$^{-1}$, Orthophosphate ≤ 0.4 mg.L$^{-1}$; further details in Tableau S2). Moreover, if habitat rarely or never undergoes sufficient immersion by alkaline water, provided by ground water emergences, alkaline conditions can be lost, and an acidification process can occur. Near the area of the study, some habitats seem to undergo this process, since a change in vegetation was observed, with a fast development of *Sphagnum* species under birch trees leading in a second step towards more acidic conditions but also an elevation of the soil with a potential of *Sphagnum*-peat forming (Reeve et al., 2000; Soudzilovskaia et al., 2010). This result could be also of importance to redefine our consideration of the suitable area, selecting the quadrats with a lower mean elevation. Our results are in accordance with those found by Jabłońska et al. (2011), who described some processes shaping vegetation patterns, including one hosting a *L. loeselii* population, the brown moss- small and slender sedge fens. In a context of alkaline fen restoration, lead on the site by the Conservatoire d'espaces naturels through the LIFE project Anthropofens, an improvement of the functioning of the peatland is expected. One of the planned actions being a sodcutting to improve the trophic conditions on the superficial peat layer, and contribute to favour targeted habitats, these results give insights to meet the improvement of the functioning with the *L. loeselii* conservation. In that case, we would recommend to sodcut at least in some part of the site, until the elevation with the highest occurrence probability of *L. loeselii* individuals.

## Conclusion

We showed that the fluctuations of the population of *L. loeselii* may be explained by factors linked to water levels (Figure S3). They have different effects on the population according to the season, since they act directly on particular stages of *L. loeselii* life cycle. This work also provides insights to improve the accuracy of the targeted soil and water level elevation, and therefore a best restoration by sodcutting and weir installation. Management of water levels all year long should then be improved to meet the manager's objectives on the site. Since the Somme Valley is a strongly anthropized valley, the water level fluctuations may be more important than in other contexts. Therefore, if this study disentangled the factors relative to water levels on the *L. loeselii* population in this context, it seems difficult to generalize to other contexts with a better habitat preservation, and less drought or flooding probabilities. If we focused on water levels and elevation, the interaction with higher vegetation may be of importance for Liparis species germination and flowering (Tsutsumi et al., 2011). Therefore, models with a larger number of observations on mowing timing could be further performed to disentangle the competition processes from the one of environmental factors.

**Acknowledgement :** We would like to thank the partners of the project LIFE Anthropofens and all the people that contributed to the field work and the site management including Matthieu James, Raoul Daubresse, Patrick Trongneux, Alexis André, Jérémy Hummel, Eric Bastien, Guillaume Gaudin, Angélique Philippe and all the employees and administrators of the CEN Hauts-de-France. We also warmly thank the Conservatoire Botanique National de Bailleul for discussion about the L. loeselii conservation as well as the people that was involved at initiating the protocol, namely Bertille Asset, Geoffroy Villejoubert, Antoine Hébert, Vincent Lévy, and Jean-Christophe Hauguel.

**Conflict of interest :** The authors declare no conflict of interest, and the funders had no role in the writing and publication processes of the manuscript.

# Bibliography

Berquer, A., 2023. Subsidence monitoring of the peat valleys of Northern France and Wallonia. Conservatoire d'espaces naturels des Hauts-de-France, Life Anthropofens 10.

Bilz, M., 2011. Liparis loeselii (Europe assessment). The IUCN Red List of Threatened Species 2011: e.T161960A5519865.

Błońska, A., Halabowski, D., Sowa, A., 2016. Population structure of *Liparis loeselii* (L.) Rich. in relation to habitat conditions in the Warta River valley (Poland). Biodiversity Research and Conservation 43, 41–52. https://doi.org/10.1515/biorc-2016-0016

Bonn, A., Allott, T., Evans, M., Joosten, H., Stoneman, R., 2016. Peatland restoration and ecosystem services: an introduction, in: Bonn, A., Allott, T., Evans, M., Joosten, H., Stoneman, R. (Eds.), Peatland Restoration and Ecosystem Services. Cambridge University Press, pp. 1–16. https://doi.org/10.1017/CBO9781139177788.002

Convention on Wetlands, 2021. Global guidelines for peatland rewetting and restoration. Ramsar Technical Report No. 11. Convention on Wetlands, Gland, Switzerland.

Destiné, B., 2000. La germination et la croissance juvénile de Liparis loeselii (L.) L.C.M. Rich. en conditions de culture asymbiotique in vitro [Exemple de sauvetage d'une espèce sauvage en voie de disparition grâce à la biotechnologie], in: Conservation ex situ des plantes menacées - Compte rendu & communications du groupe de travail, session organisée à Bailleul du 17 au 20 janvier 2000, Centre Régional de Phytosociologie / Conservatoire Botanique National de Bailleul. pp. 193–206.

François, R., 2021. Les 15 000 hectares de tourbières alcalines des vallées de Somme et d'Avre (Picardie) Première partie : milieu physique et géohistoire. Bull. Société Linnéenne Nord-Picardie 77–160.

Gabry, J., Veen, D., Stan Development Team, Andreae, M., Betancourt, M., Carpenter, B., Gao, Y., Gelman, A., Goodrich, B., Lee, D., Song, D., Trangucci, R., 2022. Package 'shinystan' : Interactive Visual and Numerical Diagnostics and Posterior Analysis for Bayesian Models. version 2.6.0. https://mc-stan.org/shinystan/.

Groot, R.S. de, Stuip, M., Finlayson, C.M., Davidson, N., 2006. Valuing wetlands: guidance for valuing the benefits derived from wetland ecosystem services. Ramsar Convention Secretariat Secretariat of the Convention on Biological Diversity, Gland, Montreal.

Grootjans, A., Shahrudin, R., van de Craats, A., Kooijman, A., Oostermeijer, G., Petersen, J., Amatirsat, D., Bland, C., Stuyfzand, P., 2017. Window of opportunity of Liparis loeselii populations during vegetation succession on the Wadden Sea islands. J Coast Conserv 21, 631–641. https://doi.org/10.1007/s11852-016-0448-6

Hatala, J.A., Detto, M., Sonnentag, O., Deverel, S.J., Verfaillie, J., Baldocchi, D.D., 2012. Greenhouse gas ($CO_2$, $CH_4$, $H_2O$) fluxes from drained and flooded agricultural peatlands in the Sacramento-San Joaquin Delta. Agriculture, Ecosystems & Environment 150, 1–18. https://doi.org/10.1016/j.agee.2012.01.009

Hendoux, F., Destiné, B., Bertrand, J., 2001. Plan de conservation du Liparis de Loesel [Liparis Loeselii (L.) L.C.M. Rich.]. DIREN Nord-Pas-de-Calais 86.

Herrera-García, G., Ezquerro, P., Tomás, R., Béjar-Pizarro, M., López-Vinielles, J., Rossi, M., Mateos, R.M., Carreón-Freyre, D., Lambert, J., Teatini, P., Cabral-Cano, E., Erkens, G., Galloway, D., Hung, W.-C., Kakar, N., Sneed, M., Tosi, L., Wang, H., Ye, S., 2021. Mapping the global threat of land subsidence. Science 371, 34–36. https://doi.org/10.1126/science.abb8549

IGN, 2018. RGE ALTI® Version 2.0 - Descriptif de contenu.

Ikkala, L., Ronkanen, A.-K., Utriainen, O., Kløve, B., Marttila, H., 2021. Peatland subsidence enhances cultivated lowland flood risk. Soil and Tillage Research 212, 105078. https://doi.org/10.1016/j.still.2021.105078

Jabłońska, E., Pawlikowski, P., Jarzombkowski, F., Chormański, J., Okruszko, T., Kłosowski, S., 2011. Importance of water level dynamics for vegetation patterns in a natural percolation mire (Rospuda fen, NE Poland). Hydrobiologia 674, 105–117. https://doi.org/10.1007/s10750-011-0735-z

Joosten, H., 2016. Peatlands across the globe, in: Bonn, A., Allott, T., Evans, M., Joosten, H., Stoneman, R. (Eds.), Peatland Restoration and Ecosystem Services. Cambridge University Press, pp. 19–43. https://doi.org/10.1017/CBO9781139177788.003

Jurasinski, G., Ahmad, S., Anadon-Rosell, A., Berendt, J., Beyer, F., Bill, R., Blume-Werry, G., Couwenberg, J., Günther, A., Joosten, H., Koebsch, F., Köhn, D., Koldrack, N., Kreyling, J., Leinweber, P., Lennartz, B., Liu, H., Michaelis, D., Mrotzek, A., Negassa, W., Schenk, S., Schmacka, F., Schwieger, S., Smiljanić, M., Tanneberger, F., Teuber, L., Urich, T., Wang, H., Weil, M., Wilmking, M., Zak, D., Wrage-Mönnig, N., 2020. From Understanding to Sustainable Use of Peatlands: The WETSCAPES Approach. Soil Syst. 4, 14. https://doi.org/10.3390/soilsystems4010014

Kotowski, W., Dzierża, P., Czerwiński, M., Kozub, Ł., Śnieg, S., 2013. Shrub removal facilitates recovery of wetland species in a rewetted fen. Journal for Nature Conservation 21, 294–308. https://doi.org/10.1016/j.jnc.2013.03.002

Kreyling, J., Tanneberger, F., Jansen, F., Van Der Linden, S., Aggenbach, C., Blüml, V., Couwenberg, J., Emsens, W.-J., Joosten, H., Klimkowska, A., Kotowski, W., Kozub, L., Lennartz, B., Liczner, Y., Liu, H., Michaelis, D., Oehmke, C., Parakenings, K., Pleyl, E., Poyda, A., Raabe, S., Röhl, M., Rücker, K., Schneider, A., Schrautzer, J., Schröder, C., Schug, F., Seeber, E., Thiel, F., Thiele, S., Tiemeyer, B., Timmermann, T., Urich, T., Van Diggelen, R., Vegelin, K., Verbruggen, E., Wilmking, M., Wrage-Mönnig, N., Wołejko, L., Zak, D., Jurasinski, G., 2021. Rewetting does not return drained fen peatlands to their old selves. Nat Commun 12, 5693. https://doi.org/10.1038/s41467-021-25619-y

Leifeld, J., Müller, M., Fuhrer, J., 2011. Peatland subsidence and carbon loss from drained temperate fens: Peatland drainage and carbon loss. Soil Use and Management 27, 170–176. https://doi.org/10.1111/j.1475-2743.2011.00327.x

Marescaux, Q., Lebrun, J., Gaudin, G., 2021. Plan régional d'action en faveur des tourbières des Hauts-de-France 2022-2031.

McInnes, R.J., 2013. Recognizing Ecosystem Services from Wetlands of International Importance: An Example from Sussex, UK. Wetlands 33, 1001–1017. https://doi.org/10.1007/s13157-013-0458-1

Megre, D., Roze, D., Dokane, K., Jakobsone, G., Karlovska, A., 2018. Survival of an Endangered Orchid *Liparis loeselii* in Habitats with Different Water Level Fluctuations. Polish Journal of Ecology 66, 126–138. https://doi.org/10.3161/15052249PJE2018.66.2.004

Meire, G., 2019. Conservation du Liparis de Loesel sur le marais tourbeux de Blangy-Tronville (Somme, Hauts-de-France) : dynamique de la population, influence des. Bull. Société Linnéenne Nord-Picardie 37, 128–138.

Meire, G., Rivière, G., 2019. APPB du Grand Marais de la Queue et autres marais communaux de Blangy- Tronville (Somme). Evaluation du plan de gestion 2012 -2016 et nouveau plan 2019-2029. Conservatoire d'espaces naturels de Picardie 220.

Middleton, B.A., Holsten, B., van Diggelen, R., 2006. Biodiversity management of fens and fen meadows by grazing, cutting and burning. Applied Vegetation Science 9, 307–316. https://doi.org/10.1111/j.1654-109X.2006.tb00680.x

Muth, C., Oravecz, Z., Gabry, J., 2018. User-friendly Bayesian regression modeling: A tutorial with rstanarm and shinystan. TQMP 14, 99–119. https://doi.org/10.20982/tqmp.14.2.p099

R Core Team, 2023. R: A Language and Environment for Statistical Computing. R Foundation for Statistical Computing.

Reeve, A.S., Siegel, D.I., Glaser, P.H., 2000. Simulating vertical flow in large peatlands. Journal of Hydrology 227, 207–217. https://doi.org/10.1016/S0022-1694(99)00183-3

Rolfsmeier, S.B., 2007. Liparis loeselii (L.) Rich. (yellow widelip orchid): a technical conservation assessment.

Roze, D., Jakobsone, G., Megre, D., Belogrudova, I., Karlovska, A., 2014. Survival of Liparis loeselii (L.) as an early successional species in Engure region described based on ecological peculiarities DURING the annual cycle. Proceedings of the Latvian Academy of Sciences. Section B. Natural, Exact, and Applied Sciences. 68, 93–100. https://doi.org/10.2478/prolas-2014-0008

Sivula, T., Magnusson, M., Matamoros, A.A., Vehtari, A., 2022. Uncertainty in Bayesian Leave-One-Out Cross-Validation Based Model Comparison.

Sonnier, G., Boughton, E.H., Whittington, R., 2023. Long-term response of wetland plant communities to management intensity, grazing abandonment, and prescribed fire. Ecological Applications 33. https://doi.org/10.1002/eap.2732

Soudzilovskaia, N.A., Cornelissen, J.H.C., During, H.J., Aerts, R., 2010. Similar cation exchange capacities among bryophyte species refute a presumed mechanism of peatland acidification. Ecology 91, 2716–2726.

Stan Development Team, 2023. Stan Modeling Language Users Guide and Reference Manual, version 2.21.4. https://mc-stan.org.

Takeno, K., 2016. Stress-induced flowering: the third category of flowering response. EXBOTJ 67, 4925–4934. https://doi.org/10.1093/jxb/erw272

Tsutsumi, C., Miyoshi, K., Yukawa, T., Kato, M., 2011. Responses of seed germination and protocorm formation to light intensity and temperature in epiphytic and terrestrial *Liparis* (Orchidaceae). Botany 89, 841–848. https://doi.org/10.1139/b11-066

Valentin, B., Toussaint, B., Duhamel, F., Valet, J.-M., 2010. Plan National d'Action : Liparis de Loesel (2010-2014). Conservatoire Botanique National de Bailleul, Ministère de l'Ecologie, de l'Energie, du Développement Durable et de la Mer.

Van Landuyt, W., T'jollyn, F., Brys, R., Vanden Broeck, A., 2015. Translocatie-experiment bij groenknolorchis (Liparis loeselii). Rapporten van het Instituut voor Natuur- en Bosonderzoek (No. NBO.R.2015.10142746)). Instituut voor Natuur- en Bosonderzoek, Brussel.

Vanden Broeck, A., Van Landuyt, W., Cox, K., De Bruyn, L., Gyselings, R., Oostermeijer, G., Valentin, B., Bozic, G., Dolinar, B., Illyés, Z., Mergeay, J., 2014. High levels of effective long-distance dispersal may blur ecotypic divergence in a rare terrestrial orchid. BMC Ecol 14, 20. https://doi.org/10.1186/1472-6785-14-20

Wheeler, B.D., Lambley, P.W., Geeson, J., 1998. Liparis loeselii (L.) Rich. in eastern England: constraints on distribution and population development. Botanical Journal of the Linnean Society 126, 141–158. https://doi.org/10.1111/j.1095-8339.1998.tb02522.x

Zak, D., McInnes, R.J., 2022. A call for refining the peatland restoration strategy in Europe. Journal of Applied Ecology 59, 2698–2704. https://doi.org/10.1111/1365-2664.14261

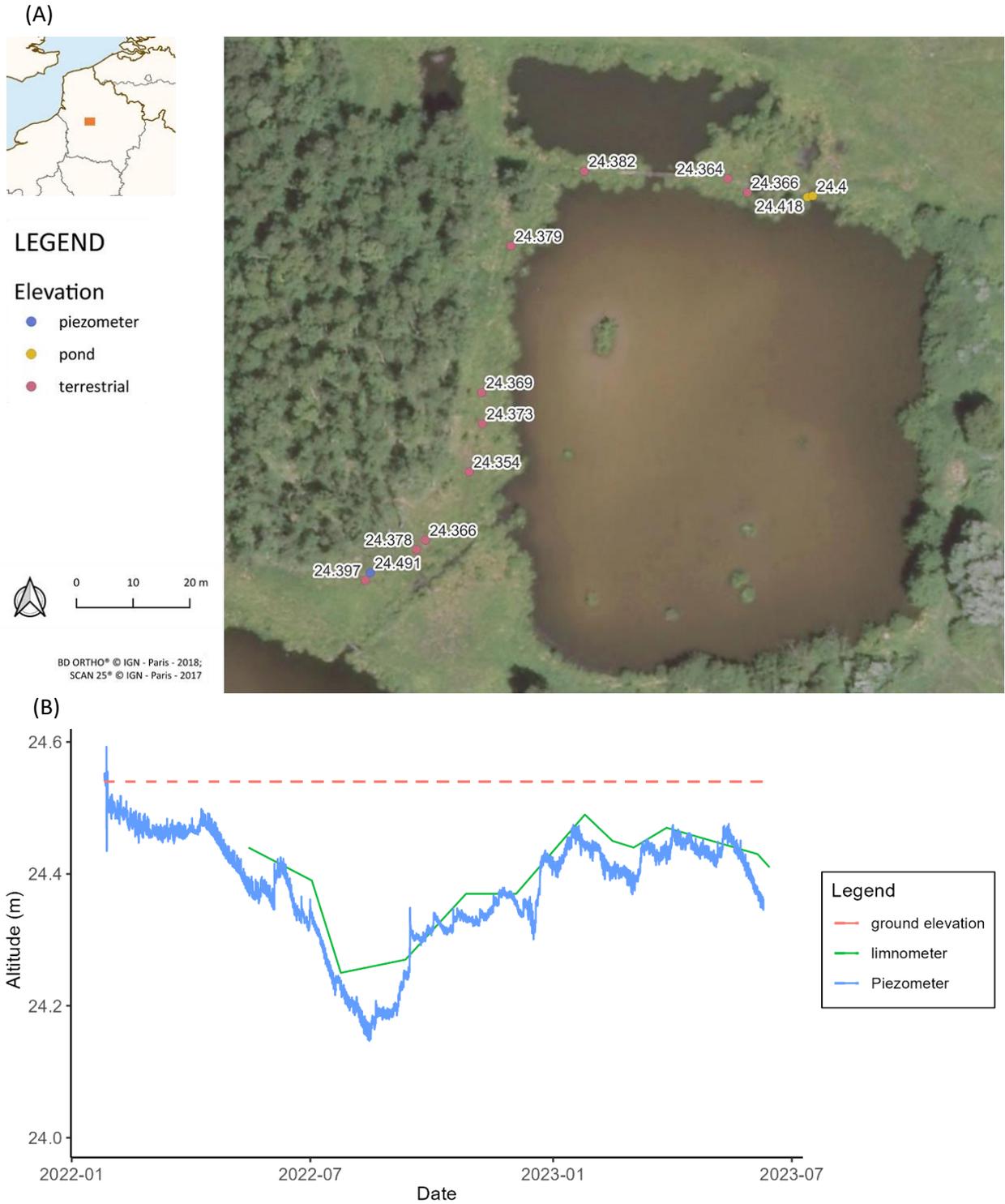

*Figure S1: Water levels measured (A) the same day (August 30th, 2023) in the pond and in microtopographic depressions on the terrestrial area covered with a thin layer of water, and (B) on the limnometric scale (green) and with the piezometer (blue) that was installed in January 2022, at a ground elevation of 24.50 m.*

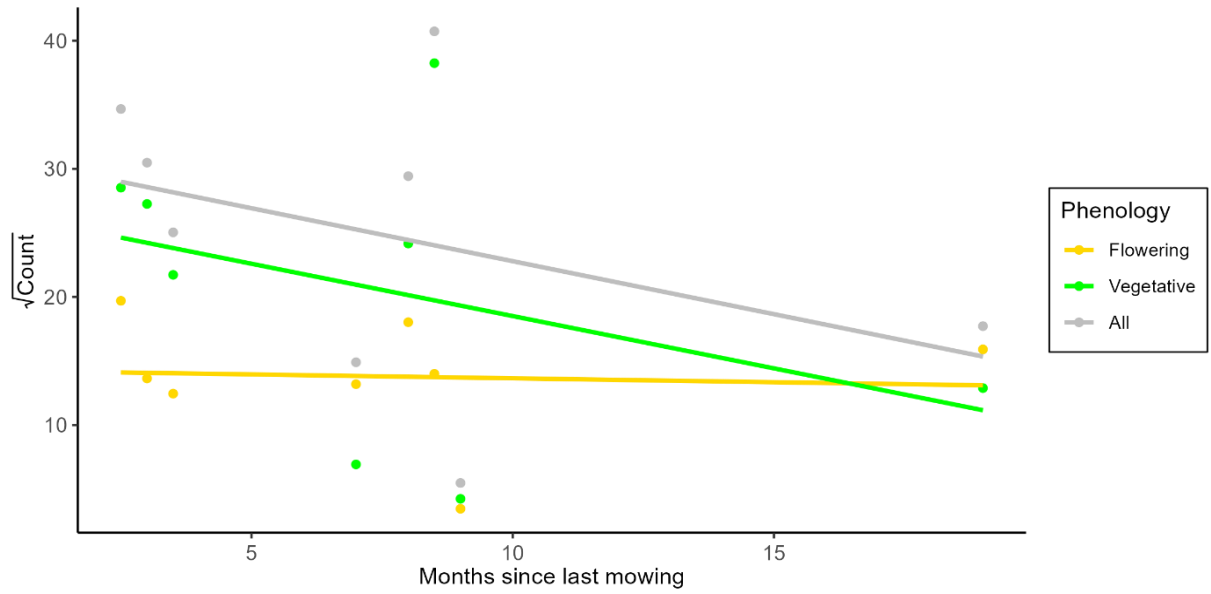

*Figure S2: Linear regression of the square-root transformed count of Liparis of each stage according to the time since the last mowing. Mowing generally takes place in autumn or in spring, when the conditions allow it : avoiding the vegetative and flowering stages of L. loeselii and avoiding the high water levels period that does not allow the mechanical use. These relationships are not significant (P-Values > 0.05, Number of observations = 8, yellow: R²(flowering) = 0.004, green: R²(vegetative) = 0.139, grey: R²(total) = 0.146).*

*Table S1: Results of the Welch t-test comparing the mean (upper table) and the standard deviation (lower table) of the elevation of quadrats including at least one individual of L. loeselii (presence) with a quadrat where no individual was reported (absence).*

| Variable | Year | Mean values of the variable for quadrat | | t-value | DF | p-value | |
|---|---|---|---|---|---|---|---|
| | | Presence | Absence | | | | |
| Mean Elevation | 2007 | 24.594 | 24.713 | -9.642 | 144 | <0.001 | *** |
| | 2008 | 24.597 | 24.711 | -8.385 | 79 | <0.001 | *** |
| | 2009 | 24.588 | 24.709 | -7.748 | 34 | <0.001 | *** |
| | 2010 | 24.608 | 24.715 | -8.055 | 161 | <0.001 | *** |
| | 2011 | 24.589 | 24.722 | -11.602 | 333 | *<0.001* | *** |
| | 2012 | 24.593 | 24.726 | -10.982 | 332 | *<0.001* | *** |
| | 2013 | 24.584 | 24.717 | -12.087 | 311 | <0.001 | *** |
| | 2014 | 24.584 | 24.721 | -12.380 | 341 | <0.001 | *** |
| | 2015 | 24.588 | 24.737 | -12.806 | 342 | <0.001 | *** |
| | 2017 | 24.593 | 24.732 | -11.747 | 355 | <0.001 | *** |
| | 2018 | 24.587 | 24.736 | -13.128 | 330 | <0.001 | *** |
| | 2019 | 24.583 | 24.732 | -13.314 | 345 | <0.001 | *** |
| | 2021 | 24.581 | 24.712 | -12.316 | 272 | <0.001 | *** |
| | 2023 | 24.584 | 24.707 | -7.514 | 15 | <0.001 | *** |
| Standard deviation | 2007 | 0.031 | 0.051 | -5.605 | 50 | <0.001 | *** |
| | 2008 | 0.034 | 0.050 | -4.649 | 40 | <0.001 | *** |
| | 2009 | 0.028 | 0.050 | -4.640 | 19 | <0.001 | *** |
| | 2010 | 0.033 | 0.051 | -5.779 | 91 | <0.001 | *** |
| | 2011 | 0.032 | 0.052 | -6.577 | 133 | <0.001 | *** |
| | 2012 | 0.032 | 0.052 | -7.156 | 184 | <0.001 | *** |
| | 2013 | 0.029 | 0.051 | -8.163 | 120 | <0.001 | *** |
| | 2014 | 0.030 | 0.052 | -7.405 | 122 | <0.001 | *** |
| | 2015 | 0.030 | 0.054 | -9.622 | 338 | <0.001 | *** |
| | 2017 | 0.033 | 0.053 | -7.682 | 293 | <0.001 | *** |
| | 2018 | 0.032 | 0.054 | -8.267 | 321 | <0.001 | *** |
| | 2019 | 0.029 | 0.054 | -9.963 | 318 | <0.001 | *** |
| | 2021 | 0.031 | 0.050 | -7.334 | 78 | *<0.001* | *** |
| | 2023 | 0.034 | 0.049 | -3.692 | 10 | 0.004 | ** |

*Tableau S2: Chemical analysis of different parameters of water samples in the ponds of the 'Grand marais de la queue' in Blangy-Tronville, Somme, France. Bla_G1 and Bla 6 are the samples the closer from the Liparis area*

| Sample | Date | Site | Session | Season | Year | Hour | pH | Dissolved oxygen_mg_L | pc_oxy | Conductivity µS_cm | Temperature_C | N Kjeldahl_mg_L | Nitrates_mg_l | Nitrites_mg_L | Ortho-phosphates_mg_L | Turbidite | TAC | Carbonates | Hydrogeno carbonates | Ca_total |
|---|---|---|---|---|---|---|---|---|---|---|---|---|---|---|---|---|---|---|---|---|
| Bla_G1 | 12/11/2020 | Blangy | S1 | automne | 2021 | 13h28 | na | na | na | 392 | 12.1 | 0.7 | 0 | 0 | 0 | 1.2 | 16.1 | 0 | 196 | 60.19 |
| Bla_G2 | 12/11/2020 | Blangy | S1 | automne | 2021 | 14h11 | na | na | na | 392 | 11 | 0.6 | 0 | 0 | 0 | 1.3 | 16.6 | 0 | 203 | 59.05 |
| Bla_P1 | 12/11/2020 | Blangy | S1 | automne | 2021 | 14h20 | na | na | na | 396 | 10.8 | 0.9 | 0 | 0 | 0 | 1.2 | 14.2 | 0 | 173 | 57.7 |
| Bla_3 | 12/11/2020 | Blangy | S1 | automne | 2021 | 13h17 | na | na | na | 393 | 11.4 | 0.7 | 0 | 0 | 0 | 1.4 | 16.7 | 0 | 204 | 57.46 |
| Bla_4 | 12/11/2020 | Blangy | S1 | automne | 2021 | 13h38 | na | na | na | 394 | 12 | 0.7 | 0 | 0 | 0 | 2.2 | 16.5 | 0 | 201 | 56.56 |
| Bla_5 | 12/11/2020 | Blangy | S1 | automne | 2021 | 13h47 | na | na | na | 394 | 11.6 | 0.8 | 0 | 0 | 0 | 1.9 | 16.4 | 0 | 200 | 57.56 |
| Bla_6 | 12/11/2020 | Blangy | S1 | automne | 2021 | 13h59 | na | na | na | 394 | 11.4 | 0.7 | 0 | 0 | 0 | 1.1 | 16.2 | 0 | 196 | 58.19 |
| Bla_G1 | 06/05/2021 | Blangy | S2 | printemps | 2021 | 12h50 | 7.95 | 6.15 | 57.8 | 499 | 12 | 0.7 | 0 | 0 | 0 | 1.8 | 22.6 | 0 | 276 | 79.95 |
| Bla_G2 | 06/05/2021 | Blangy | S2 | printemps | 2021 | 12h40 | 7.94 | 6.8 | 66.4 | 499 | 13.3 | 0.7 | 0.3 | 0 | 0 | 1.9 | 22.6 | 0 | 276 | 80.44 |
| Bla_P1 | 06/05/2021 | Blangy | S2 | printemps | 2021 | 12h25 | 7.94 | 6.91 | 65.9 | 503 | 13 | 0.7 | 0.3 | 0 | 0 | 3.5 | 22.9 | 0 | 279 | 79.93 |
| Bla_4 | 06/05/2021 | Blangy | S2 | printemps | 2021 | 13h08 | 7.92 | 6.56 | 62.3 | 496 | 12.4 | 0.7 | 0 | 0 | 0 | 2 | 22.7 | 0 | 277 | 80.09 |
| Bla_6 | 06/05/2021 | Blangy | S2 | printemps | 2021 | 13h25 | 7.97 | 6.01 | 57.5 | 497 | 12.9 | 0.6 | 0.2 | 0 | 0 | 1.4 | 22.9 | 0 | 279 | 79.12 |
| Bla_G1 | 09/11/2021 | Blangy | S3 | automne | 2021 | 12h49 | 7.85 | 6.69 | 57.7 | 389 | 8.8 | 0.6 | 0.2 | <0,05 | <0,4 | 1.7 | 16.4 | 0 | 200 | 57.62 |
| Bla_G2 | 09/11/2021 | Blangy | S3 | automne | 2021 | 12h30 | 7.85 | 5.19 | 43.7 | 400 | 8.2 | 0.7 | 0.4 | <0,05 | <0,4 | 1.8 | 16.6 | 0 | 203 | 58.62 |
| Bla_P1 | 09/11/2021 | Blangy | S3 | automne | 2021 | 12h15 | 7.82 | 4.35 | 36.4 | 396 | 8 | 0.8 | 0.3 | <0,05 | <0,4 | 2 | 16.8 | 0 | 205 | 58.83 |
| Bla_4 | 09/11/2021 | Blangy | S3 | automne | 2021 | 13h00 | 7.68 | 6.14 | 53.8 | 394 | 9.7 | 0.7 | 0.1 | <0,05 | <0,4 | 4.2 | 16.4 | 0 | 200 | 58.18 |
| Bla_6 | 09/11/2021 | Blangy | S3 | automne | 2021 | 13h17 | 7.77 | 5.75 | 49.5 | 391 | 9 | 0.6 | 0.5 | <0,05 | <0,4 | 1.5 | 16.6 | 0 | 203 | 59.94 |
| Bla_G1 | 12/05/2022 | Blangy | S4 | printemps | 2022 | 12h15 | 7.9 | 3.76 | 41.7 | 486 | 20.7 | 1.1 | 0.3 | <0,05 | <0,4 | 12 | 21.6 | 0 | 263.5 | 75.47 |
| Bla_G2 | 12/05/2022 | Blangy | S4 | printemps | 2022 | 12h27 | 7.8 | 3.77 | 42.3 | 488 | 21.7 | 0.9 | <0,1 | <0,05 | <0,4 | 4.4 | 21.4 | 0 | 261 | 75.35 |
| Bla_P1 | 12/05/2022 | Blangy | S4 | printemps | 2022 | 12h39 | 7.76 | 2.98 | 34.4 | 490 | 22.4 | 1.1 | 0.2 | <0,05 | <0,4 | 2.6 | 21.6 | 0 | 263.5 | 74.88 |
| Bla_4 | 12/05/2022 | Blangy | S4 | printemps | 2022 | 12h50 | 7.88 | 3.58 | 40 | 485 | 20.5 | 1 | 0.1 | <0,05 | <0,4 | 4.6 | 20.4 | 0 | 249 | 81.32 |
| Bla_6 | 12/05/2022 | Blangy | S4 | printemps | 2022 | 13h05 | 7.92 | 3.56 | 39.5 | 492 | 20.6 | 1 | 0.2 | <0,05 | <0,4 | 3.9 | 21.8 | 0 | 266 | 81.34 |

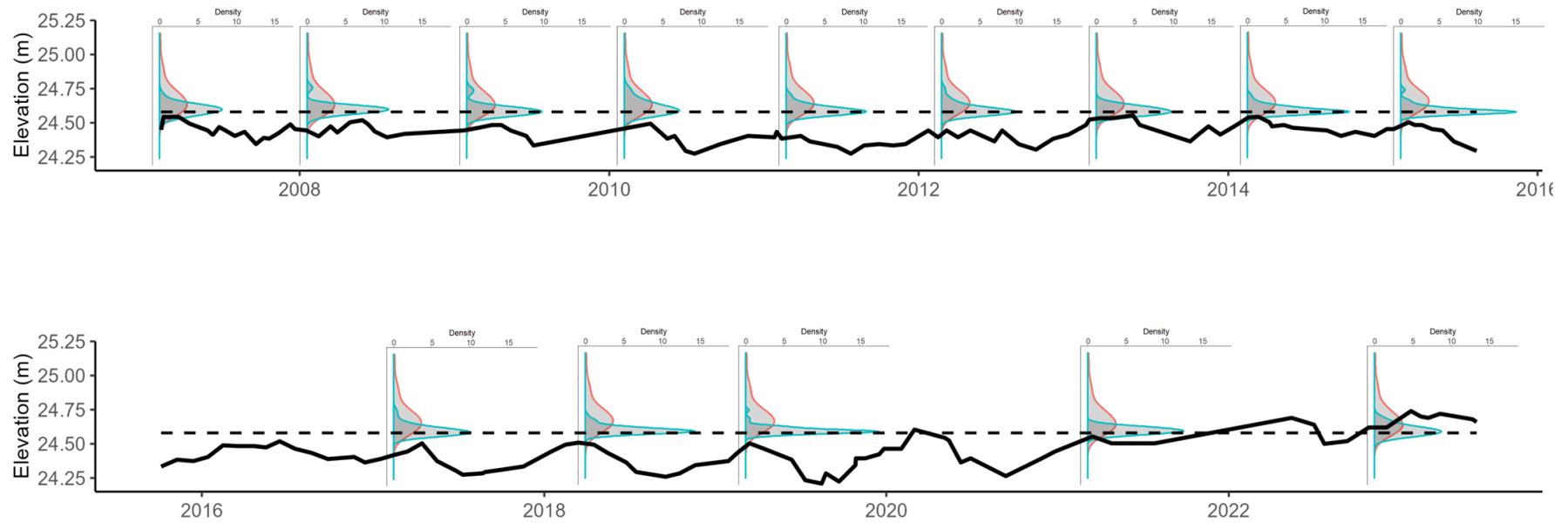

*Figure S3: Representation of the water level measured on the limnometric scale converted in elevation (m.a.s.l., black line). Each year when the quadrat protocol was performed, the distribution of the pseudo-absence quadrats (red) and presence quadrats (blue) were represented. The scale of the 'density' axis (top) is fixed, to represent the number of quadrats of each stage per year as well as the variance of quadrat hosting L. loeselii individuals. The black dashed line represents the mean ground level of quadrats where at least one individual was observed, around 24.53 m.a.s.l.. The x-axis covers the whole time span of the study, from 2007 to 2023.*